\documentclass{ws-procs9x6}
\usepackage{epsfig}

\begin{document}

\title{\flushright{TPJU-17/2004}\\
Exotic Challenges\footnote{\uppercase{T}his work is supported
by the \uppercase{P}olish \uppercase{S}tate \uppercase{C}ommittee
for \uppercase{S}cientific \uppercase{R}esearch (\uppercase{KBN})
under grant 2 \uppercase{P}03\uppercase{B} 043 24.
\uppercase{T}alk at \uppercase{I}nternational \uppercase{W}orkshop
\uppercase{PENTAQUARK}04, \uppercase{S}pring-8, \uppercase{J}uly
20--23, 2004 }}

\author{M. PRASZA{\L}OWICZ}

\address{M. Smoluchowski Institute of Physics, \\
Jagellonian University, \\
ul. Reymonta 4, 30-059 Krak{\'o}w, Poland\\
E-mail: michal@if.uj.edu.pl}

\maketitle

\abstracts{We list and discuss theoretical consequences of recent
discovery of $\Theta^+$. }

\section{Has $\Theta^+$ been really found?} \label{s1}

Let us start with a word of warning. No evidence of $\Theta^+$ has
been found in HERA-B ~\cite{Knopfle:2004tu},
RHIC~\cite{RHIC}, BES~\cite{Bai:2004gk},
LEP~\cite{LEP} and
Fermilab~\cite{FNAL}.
The reasons maybe either of experimental nature or a peculiar
production mechanism~\cite{Karliner:2004gr,Lipkin:2004trans}.

In contrast to the low energy almost fully exclusive experiments
that reported $\Theta^{+}$, experiments which do not see exotics
are mostly high energy inclusive ones~\cite{Hicks:2004vd}. It is
difficult to produce exotic states in the high energy experiments
which are dominated by the Pomeron exchanges
\cite{Diakonov:2004ie,Titov:2004trans}. Note that experiments
which do not see $\Theta^{+}$ put in fact an upper bound on the
(yet unknown) production mechanism, rather than exclude its
existence.

Another piece of negative evidence comes from the old $KN$
scattering data that have been recently
reanalyzed~\cite{Arndt,Gibbs:2004ji}. Here one can accommodate at
most one resonance near 1545 MeV with very small width
$\Gamma_{\Theta^{+}}<2$ MeV. $K^{+}d$ cross-section including the
hypothesis of a narrow resonance recalculated in the J\"{u}lich
meson exchange model~\cite{Sibirtsev:2004bg} yields
$\Gamma_{\Theta^{+}}<1$ MeV. However, ''non-standard'' analysis of
the phase shifts allows for more exotics~\cite{Kelkar,Kelkar1}.

All these facts call for a new high precision $KN$ experiment in the
interesting energy range.

\section{How many $\Theta$'s~?}

\begin{figure}[h]
\centerline{\epsfxsize=2.5in\epsfbox{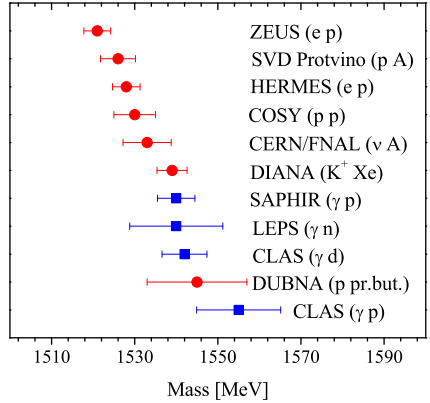}}
\caption{Mass of $\Theta^{+}$ as reported by various experiments. Statistical
and stystematic errors have been added in quadrature. Squares refer to $K^{+}
n$ final state and circles to $K^{0}_{s} p$. }%
\label{fig:data}%
\end{figure}

Since the first report on $\Theta^+$ by LEPS~\cite{Nakano:2003qx}
many other experiments confirmed its existence~\cite{Thetaexp}.
Reported masses are shown in Fig.~\ref{fig:data}. Some of these
results were reported at this Workshop~\cite{Penta04} together
with new results from LEPS~\cite{Nakano:2004trans}. In principle
data in Fig.~\ref{fig:data} should represent one state. However,
if taken literary, ZEUS and CLAS data for example are not
compatible.

\begin{figure}[th]
\centerline{\epsfxsize=2.8in\epsfbox{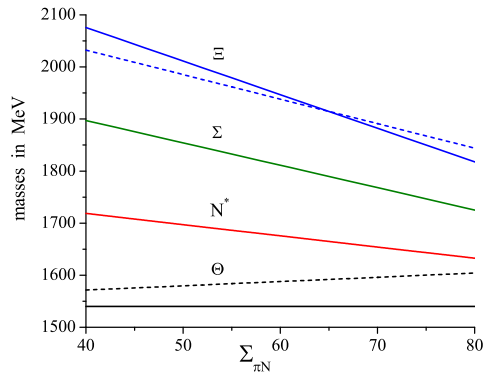}}
\caption{The spectra of $\overline{10}_{1/2}$ baryons (solid lines) together
with the masses of the $\Theta_{1}$ and $\Xi_{3/2}$ in the $27_{3/2}$ (dashed
lines) as functions of $\Sigma_{\pi N}$, using parameters fitted from the
masses of the $\Theta^{+}$ and non-exotic states. }%
\label{fig:freedom}%
\end{figure}

It is therefore legitimate to ask: do all these experiments see
the same state? Before this issue is decided experimentally let us
examine predictions of different models. Chiral models predict a
tower of exotic rotational states starting with $\overline
{10}_{1/2}$, $27_{3/2,1/2}$, $\overline{35}_{5/2,3/2}$ (subscripts
refer to spin) etc. The lowest excitation of $\Theta^{+}$ is an
isospin triplet of spin 3/2 belonging to flavor $27$. The mass
$\Theta_{27}$ is only slightly larger than the mass of
$\Theta^{+}$ and depends weakly on the value of pion nucleon
$\Sigma_{\pi N}$ term (see Fig.~\ref{fig:freedom}). Note that
theoretical uncertainty of the model~\cite{Ellis:2004uz} is
approximately $\pm30$ MeV.

In the correlated quark models additional states are also
unavoidable. In the diquark model~\cite{Jaffe:2003sg} the
spin-orbit interaction splits spin 1/2 and 3/2 states by a tiny
amount of a $\Delta E\sim35\div65$ MeV~\cite{Dudek:2003xd}.
Similarly in the diqaurk-triquark scenario~\cite{Karliner:2003dt},
the mass splitting would be of the order of $40$ MeV. Hence a
nearby isosinglet $\Theta^{\ast}$ state of spin $3/2$ is expected
in these models. This is a distinguishing feature, since the
soliton models do not accommodate spin $3/2$ antidecuplet.

Although there are no more exotics in the minimal diquark
model~\cite{Jaffe:2003sg}, the tensor diquarks in $6$ of SU(3)
flavor are almost unavoidable. They lead to further exotics like
$27$ which in the schematic model of Shuryak and
Zahed~\cite{Shuryak} is even lighter than
$\overline{10}$.

We see therefore the importance of experimental searches both for
the isospin partners of $\Theta^{+}$ and for another peak in the
$\Theta^{+}$ channel. Preliminary CLAS
results~\cite{deVita:2004trans} indicate two states at 1523 and
1573 MeV, similarly bubble chamber experiment analyzed by the
Yerevan group~\cite{Thetaexp} reports 3 states at 1545, 1612 and
1821 MeV. Finaly, there is also report of a number of exotic
resonanses from Dubna~\cite{Thetaexp} and from the
''non-standard'' phase shift analysis~\cite{Kelkar1}. So far the
searches for $\Theta^{++}$ provided no
evidence~\cite{Thetaexp,Juengst:2003yy,Chekanov:2004hd} although
some structures in $K^{+}p$ channel have been seen by
CLAS~\cite{Juengst:2003yy} and STAR~\cite{Kabana:2004trans}. There
is no evidence for $\Theta^{++}$ in the old $K^{+}p$ scattering
data~\cite{Arndt}.

\section{Spin and parity of $\Theta^+$}

Spin and parity of $\Theta^+$ are at present unknown. While almost
all theorists agree that spin should be 1/2 the parity
distinguishes between different models. Chiral models predict
positive parity, similarly quark models with flavor dependent
forces and correlated quark models predict ${\mathcal P}=+$. In
uncorrelated quark models and sum rules ${\mathcal P}=-$.

Unlike model calculations lattice simulations (summarized at this
Workshop by S. Sasaki~\cite{Sasaki:2004trans}) should give clean
theoretical answer whether pentaquarks exists and what their
quantum numbers are. However, since pentaquarks are excited QCD
states, lattice simulations are difficult and give ambiguous
message: either there is no bound $\Theta^+$ state
\cite{noTheta}, or there is one but with negative
parity~\cite{negTheta}. One
simulation indicates~\cite{Chiu:2004gg} ${\mathcal P}=+$.

Let us stress that, unlike in the case of $\Omega^-$ whose spin
and parity are not measured but {\em assumed} after the quark
model~\cite{PDG}, the parity of $\Theta^+$ is of utmost importance
to discriminate between various models and to understand how QCD
binds quarks.

\section{The width of $\Theta^+$}

A key prediction of the seminal paper by Diakonov, Petrov and
Polyakov~\cite{Diakonov:1997mm} (DPP) was the observation that (in
contrast to the naive expectations) in the chiral quark soliton
model antidecuplet states should be very narrow. The decay width
for
$B\rightarrow B^{\prime}+\varphi$ is given by:%
\begin{equation}
\Gamma_{B\rightarrow B^{\prime}+\varphi}=\frac{G_{R}^{2}}{8\pi}\frac{p^{3}%
}{M\,M^{\prime}}\,C(B^{\prime},B,\varphi). \label{Gamma}%
\end{equation}
Here $M$ and $M^{\prime}$ are baryon masses, $p$ is meson momentum
in the $B$ rest frame, $C$ denotes pertinent SU(3) Clebsch-Gordan
coefficient and $G_{R}$ stands for a coupling constant for baryon
$B$ in the SU(3) representation $R$. It has been
observed~\cite{Diakonov:1997mm} that $G_{\overline{10}}\equiv0$ in
the nonrelativistic limit of the soliton model which is very
useful as a first approximation. This was a clear indication that
$\overline{10}$ baryons would be narrow. How narrow is of course a
question of reliability of approximations employed to derive
(\ref{Gamma}) and the phenomenological input used to determine
$G_{\overline{10}}$. DPP~\cite{Diakonov:1997mm} made a
conservative estimate that $\Gamma_{\Theta^{+}}<15$ MeV. In a more
recent analysis they have argued that
$\Gamma_{\Theta^{+}}\sim3.6\div11.2$ MeV \cite{Diakonov:2004ai}.

In the diquark models $\Theta^{+}$ decay proceeds via diquark
breakup and is therefore believed to be small. Recently it was
shown~\cite{Carlson:2003xb} that the narrowness of $\Theta^{+}$ in
the quark model with flavor-spin interactions follows from the
group-theoretical structure of the wave function.

\begin{figure}[th]
\centerline{\epsfig{file=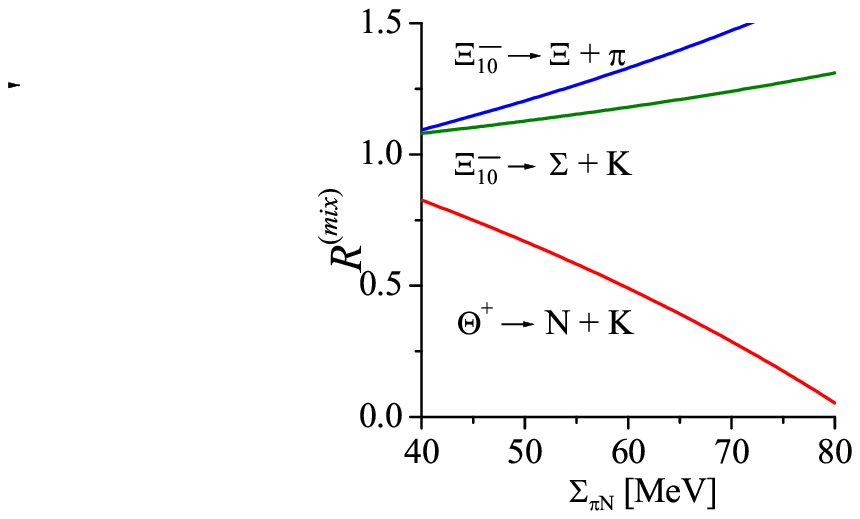,width=1.8in,clip=}}
\caption{Correction coefficients $R^{(\mathrm{mix})}$ for ${\Theta}^{+}$ and
${\Xi}_{\overline{10}}$ decays as functions of ${\Sigma}_{\pi N}$. Large
supression of $\Theta^{+}$ together with a moderate enhancement of
$\Xi_{\overline{10}}$ leads to strong SU(3) violation for the decay widths.}%
\label{fig:RTh}%
\end{figure}

Further suppression comes from the SU(3) breaking corrections due
the mixing with other representations for
$m_{s}\neq0$~\cite{Ellis:2004uz,Praszalowicz:2004dn}. Therefore
moderate admixtures of other representations for which the
relevant couplings are not suppressed may substantially modify the
decay width. In the case of $\Theta^{+}\rightarrow KN$ the
admixtures of $\overline{10}$ and $27$ in the wave function of the
final nucleon affect the decay width. In the quark-soliton model
they further suppress  $\Gamma_{\Theta^+}$ by a factor of $0.2$
\cite{Ellis:2004uz,Praszalowicz:2004dn}.  In Fig.~\ref{fig:RTh} we
show modification factor $R^{(mix)}$ for the width of $\Theta^{+}$
and for two partial widths of $\Xi_{\overline {10}}$ coming from
representation mixing in the chiral quark-soliton
model~\cite{Ellis:2004uz,Praszalowicz:2004dn} as functions of the
pion-nucleon $\Sigma_{\pi N}$ term. To conclude: the decay widths
within the antidecuplet may substantially differ from the SU(3)
symmetry values.

On experimental side the results for $\Theta^{+}$ width are
unclear. Most experiments quote upper limits, however there are a
few which claim to have measured $\Gamma_{\Theta^{+}}$ and quote
error bars. ZEUS gives~\cite{Thetaexp}: $
\Gamma_{\Theta^{+}}=6.1\pm1.6\pm ^{2.0}_{1.6}\,\quad\text{MeV}%
$ This result is consistent with the upper limit from DIANA
($K^{+}+Xe$): $\Gamma_{\Theta^{+}}<9$ MeV \cite{Thetaexp}.
Results from a $C_{3}H_{8}$ bubble chamber in
Dubna by the Yerevan group~\cite{Thetaexp}: $
\Gamma_{\Theta^{+}}=16.3\pm3.6\quad\text{MeV}%
$, from COSY~\cite{Thetaexp}: $
\Gamma_{\Theta^{+}}=18\pm4\quad\text{MeV}%
$ and Hermes~\cite{Thetaexp}: $
\Gamma_{\Theta^{+}}=19\pm5\,(\text{stat})\pm2\,(\text{syst})\quad\text{MeV}%
$ are two times larger. As discussed in Sect.~\ref{s1} old $K$
scattering data put the lowest limit $\Gamma_{\Theta^{+}}< 1 \div
2$~MeV.

In almost all theoretical models mechanisms were found which
suppress $\Theta^+$ decay width. The question is now: how much?
Therefore the measurement of the $\Theta^+$ width is of utmost
importance and will provide constraints on various theoretical
scenarios.

\section{Exotic cascades}

So far only one experiment~\cite{Alt:2003vb} reported the states
which form the "base" of $\overline{10}$, namely $I=3/2$
$\Xi^{--}_{\overline{10}}$ and $\Xi^0_{\overline{10}}$ at 1862
MeV. This result needs confirmation, so far reports from other
groups are negative.

In the original paper DPP~\cite{Diakonov:1997mm} predicted the
mass of the exotic $\Xi_{\overline{10}}$ states above 2 GeV. This
prediction, however, depends on the residual freedom of the model
which is conveniently parameterized in terms $\Sigma_{\pi
N}$~\cite{Praszalowicz:2003ik}. They used $\Sigma_{\pi N}=45$~MeV,
while present estimates~\cite{Sigma} indicate a larger value of
approximately 70~MeV. As seen from Fig.~\ref{fig:freedom} larger
values of $\Sigma_{\pi N}$ are compatible with the NA49 result.

Original prediction of the diquark model~\cite{Jaffe:2003sg} was
1750 MeV. Pure SU(3) arguments indicate that for ideal mixing
scenario employed by Jaffe and Wilczek it is difficult accommodate
exotic cascades at 1862 MeV without invoking new nucleon-like
narrow resonances~\cite{Cohen:2004gu,Praszalowicz:2004xh}. Similar
conclusion has been reached for arbitrary
mixing~\cite{Diakonov:2003jj,Pakvasa:2004pg}.

Similarly to $\Theta^+$, the decay widths of exotic cascades will
be modified by additional mixing, as depicted in
Fig.~\ref{fig:RTh}.

\section{Cryptoexotic states and mixing}

If $\Theta^{+}$ mass is 1539 MeV and $\Xi_{\overline{10}}$ 1862
MeV then equal spacing within the antidecuplet requires additional
cryptoexotic nucleon-like and $\Sigma$-like states with masses
1648 MeV and $1757$ MeV respectively. These states should be in
principle narrow with the decay widths related to
$\Gamma_{\Theta^{+}}$ by the SU(3) symmetry. However, as discussed
above, mixing will modify these relations. The nucleon-like and
$\Sigma$-like states can mix with known (and unknown) resonances
of the same parity and spin. Most of analysis in this direction
was done for two nucleon-like states $\left| S_{1,2}\right\rangle
$ assuming $J^{P}=1/2^{+}$ for antidecuplet. Here three possible
scenarios are discussed: $1)$ both states $\left|
S_{1,2}\right\rangle $ correspond to known resonances, $2)$ one
state corresponds to the yet undiscovered resonance and $3)$ both
have to be discovered.

Mixing has been also discussed by
Weigel~\cite{Weigel} within the
framework of the Skyrme model (with the dilaton field) . In this
approach, apart from rotations, another mode, namely the
''breathing'' mode of the soliton, was quantized and a subsequent
mixing with other states was investigated. Radially excited octet
states were identified with known $N^{\ast}$ resonances (Roper or
$N^{\ast}(1710)$, etc.), so that no novel states were predicted.
Unfortunately little can be said about the decay widths within
this approach.

Cohen~\cite{Cohen:2004gu} made an important remark that not only
masses but also decay widths are affected by mixing and any
phenomenological analysis should discuss both simultaneously. He
excluded ideal mixing scenario, unless new cryptoexotic
nucleon-like resonances exist.

The analysis of masses and decay widths of the $N^{\ast}$ states
under the assumption that they correspond to the Roper and
$N^{\ast}(1710)$ indicates~\cite{Pakvasa:2004pg} that it is
impossible to match the mass splittings with the observed
branching ratios even for arbitrary mixing. It is shown that the
mixing required for the decay $N^{\ast}(1710)\rightarrow
\Delta\pi$ is not compatible with the mixing deduced from the
masses.  A possibility based on the nonideal mixing scenario
advocated by Diakonov and Petrov \cite{Diakonov:2003jj} is that
there should be a new $N^{\ast}$ resonance in the mass range of
$1650\div1680$ MeV.

Similar conclusion has been reached in the quark soliton
model~\cite{Arndt:2003ga}. Here already the ordinary nucleon state
has a non-negligible admixture of $\overline{10}$ which leads to
the suppression of the decay width. Further decrease may be
achieved by adding a mixing to another nucleon-like state as Roper
and/or $N^{\ast}(1710)$ and by the admixture of
$27$~\cite{Praszalowicz:2004dn}.

The same authors \cite{Arndt:2003ga} claim that the improved phase
shift analysis admits two candidates for the narrow nucleon-like
resonances at $1680$ and $1730$ MeV and with widths smaller than
$0.5$ and $0.3$ MeV, respectively.

To conclude this Section let us note that physics of $N^{\ast}$
and $\Sigma ^{\ast}$ states will be most probably dominated by
extensive mixing between different nearby states which will affect
{\em both} masses and decay widths. New, narrow resonances are
theoretically expected. Experimental searches for such states have
been recently performed with positive preliminary evidence
\cite{Kabana:2004trans,Kouznetsov:2004trans}.

\section{Summary}

A convincing experiment confirming $\Theta^+$ is in our opinion
still missing. If $\Theta^+$ exists we have to understand why some
experiments do not see it while the others do. Although yet
unknown production mechanism might provide an explanation, it is
really hard to understand why similar experiments like ZEUS and H1
give contradictory results.

There is a common agreement that spin of $\Theta^+$ is 1/2.
However, there is no such consensus as far as parity is concerned.
Measuring the parity will discriminate between different models.
Even more importantly it will either strengthen our confidence in
lattice QCD simulations or pinpoint some yet unknown weaknesses of
this approach.

Certainly the measurement of the width is badly needed. An
intuitive explanation why $\Theta^+$ is so narrow is still missing
although in various models  formal arguments have been given.
Since the leading decay mode $\overline{10} \rightarrow 8 + 8$
(where the second 8 refers to the outgoing meson) is very small
even moderate admixtures of other SU(3) representations in the
final state or in the initial state for cryptoexotic members of
$\overline{10}$ are going to modify substantially the decay
widths. Warning: SU(3) relations between different decay widths
will not hold!

Mixing will be very important for cryptoexotic nucleon-like and
$\Sigma$-like states. Most probably, new narrow resonances are
required for consistent theoretical picture. Also the confirmation
of $\Xi_{\overline{10}}(1862)$ is badly needed.

Somewhat unexpectedly the discovery of $\Theta^+$ and possibly of
$\Xi_{\overline{10}}$ has shaken our understanding of the QCD
bound state. Simple quark model pictures must be modified and very
likely soliton models might contain necessary ingredients to
explain new exotics.

%
%
%
%

\end{document}